\documentclass{article}

\usepackage[final]{neurips_2024_ml4ps}

\usepackage[utf8]{inputenc} % allow utf-8 input
\usepackage[T1]{fontenc}    % use 8-bit T1 fonts
\usepackage{hyperref}       % hyperlinks
\usepackage{url}            % simple URL typesetting
\usepackage{booktabs}       % professional-quality tables
\usepackage{amsfonts}       % blackboard math symbols
\usepackage{nicefrac}       % compact symbols for 1/2, etc.
\usepackage{microtype}      % microtypography
\usepackage{xcolor}         % colors
\usepackage[pdftex]{graphicx}
\usepackage{placeins}

\usepackage{etoolbox}

\title{3D-PDR Orion dataset and NeuralPDR: Neural Differential Equations for Photodissociation Regions}

\author{
  Gijs Vermariën\thanks{vermarien@strw.leidenuniv.nl},~ Rahul Ravichandran \& \\ \textbf{Serena Viti}\\ \\ 
  Leiden Observatory\\
  Leiden University\\
  PO Box 9513, 2300 RA Leiden, The Netherlands \\
  \And
  Thomas G. Bisbas \\ \\
  Research Center for \\ Astronomical Computing \\
  Zhejiang Lab\\
  Hangzhou 311100, China\\
}

\newcommand{\av}{\ensuremath{A_V}}
\newcommand{\tgas}{\ensuremath{T_\mathrm{gas}}}
\newcommand{\tdust}{\ensuremath{T_\mathrm{dust}}
\newcommand{\guv}{\ensuremath{G_\mathrm{UV}}}}
\newcommand{\guvinit}{\ensuremath{G_{\mathrm{UV},0}}}
\newcommand{\densinit}{\ensuremath{n_\mathrm{H,0}}}
\newcommand{\zetainit}{\ensuremath{\zeta_0}}

\newcommand{\electron}{e$^-$}
\newcommand{\Hatom}{H}
\newcommand{\hydrogen}{H$_2$}
\newcommand{\Oatom}{O}
\newcommand{\CO}{CO}

\newcommand{\water}{H$_2$O}
\newcommand{\CH}{CH}
\newcommand{\oxygen}{O$_2$}
\newcommand{\CN}{CN}
\newcommand{\HCOp}{HCO$^+$}
\newcommand{\NH}{NH}
\newcommand{\HCN}{HCN}
\newcommand{\carbon}{C$_2$}
\newcommand{\HCO}{HCO}
\newcommand{\formaldehyde}{H$_2$CO}
\newcommand{\COp}{CO$^+$}
\newcommand{\CS}{CS}
\newcommand{\methanol}{CH$_3$OH}

\begin{document}

\maketitle

\begin{abstract}

    We present a novel dataset of simulations
    of the photodissociation region (PDR) in the Orion Bar and provide benchmarks of emulators for the dataset. 
    Numerical models of PDRs are computationally expensive since
    the modeling of these changing regions requires resolving the thermal balance and chemical composition
    along a line-of-sight into an interstellar cloud. This often makes it a bottleneck
    for 3D simulations of these regions. In this work, we provide a dataset of 8192
    models with different initial conditions  simulated with 3D-PDR. 
    We then benchmark different architectures, 
    focusing on Augmented Neural Ordinary Differential Equation (ANODE) based models \footnote{These models can be found at \url{https://github.com/uclchem/neuralpdr}.
}. Obtaining
    fast and robust emulators that can  
    be included as preconditioners of classical codes or full
    emulators into 3D simulations of PDRs.
\end{abstract}

\section{Introduction}
At the edges  of interstellar clouds there exist PDR regions that 
are dominated by UV photon chemistry. Computational models of these regions 
are expensive, since both the physical conditions and chemical composition
of these regions change as we move deeper into the cloud.
In order to simulate this accurately, we need an iterative method that 
solves for heating, cooling and chemistry, creating a visual extinction
based problem. The visual extinction is a measure for the decrease in radiation as we move into an astronomical object, and is related to the amount of hydrogen along a line of sight \citep{guverRelationOpticalExtinction2009}. By solving the differential equations describing these processes along a line of sight, we obtain the temperatures and abundances.
Having to solve along lines of sight makes comprehensive 3D simulations of these regions computationally expensive,
raising the need for surrogates that provide approximations of the
chemistry, freeing up budget for hydrodynamics and radiative transport.

In this article we provide a dataset inspired by the Orion Bar, a region within the Orion nebulae, that is currently being thoroughly investigated with the James Webb Space Telescope \citep{andree-labschModellingClumpyPhotondominated2017, habartHighangularresolutionNIRView2023a, peetersPDRs4AllIIIJWSTs2024}. 
The dataset is thus derived from typical
conditions in the Orion Bar, with radiation fields
$10^1 \leq G_\mathrm{UV}(\mathrm{Draine}) \leq 10^4$,
number densities $10^2 \leq n_\mathrm{H}(\mathrm{cm}^{-3}) \leq 10^7$ and lastly the
cosmic ray ionisations $10^{-17} \leq \zeta(\mathrm{s}^{-1})\leq 10^{-15}$. We use these parameters as initial condition for the simulation with 3D-PDR, the resulting
8192 models together we call the 3D-PDR Orion dataset\footnote{
The 3D-PDR Orion dataset is available at  \url{https://doi.org/10.5281/zenodo.13711174}}.\citep{vermarien3DPDROrionInspired2024}.

So far several machine learning based methods have been proposed to tackle 
problems in astrochemistry. Emulators have been trained to predict equilibrium chemistry \citep{demijollaIncorporatingAstrochemistryMolecular2019},
time series based chemistry \citep{grassiReducingComplexityChemical2021, holdshipChemulatorFastAccurate2021, tangReducedOrderModel2022, brancaNeuralNetworksSolving2022, sulzerSpeedingAstrochemicalReaction2023, brancaEmulatingInterstellarMedium2024} and 
position based chemistry \citep{maesMACEMachineLearning2024}. But most of these have limited accuracy,
only generalize to a small part of the physical parameter space or fail to include the interaction between
temperature and chemistry. 

We then focus on benchmarking surrogate models (emulator) that work for large physical parameter spaces
and are effective at emulating a selection of molecules that have complex formation pathways, 
without having to include the entire chemical network. To this end, we use
Augmented Neural Ordinary equations (ANODE), where the augmented part of the
NODE are auxiliary features that relate to the physical conditions of 3D-PDR models.

\section{Methods}
\subsection{Augmented Neural Ordinary Differential Equations with auxiliary parameters}
The concept of the Neural Ordinary Differential Equations \citep{chenNeuralOrdinaryDifferential2019a,kidgerNeuralDifferentialEquations2022} is that instead of defining the right hand side (RHS) of an ordinary differential equation with expert chemistry and physics knowledge, we instead 
define an approximator $\tilde{f}$. In this case we use a neural network, that 
uses data to generate a nonlinear RHS that can approximate the series. This provides an integral of the form:
\begin{equation}
y(x_2) = y(x_1) + \int_{x_1}^{x_2} \tilde{f}(x,y,e) \mathrm{d}x
\end{equation}
where $x$ is the independent variable, $y$ the dependent variables and $e$ auxiliary
parameters. 
The addition of auxiliary parameters $e$, allows us to train a model that generalizes
over many different physical models with different physical parameters. 
The usage of parameters to find more expressive NeuralODEs has been coined as
augmented ODEs \citep{dupontAugmentedNeuralODEs2019} and parameterized ODEs \citep{leeParameterizedNeuralOrdinary},
in this article we employ the term 
``auxiliary parameters'' distinguish them from the physical parameters of the  dataset and prevent confusion. Consequently, the acronym of the employed
architecture, ANODE, still fits.

These neural differential equations can be combined with encoder and decoder models \citep{kramerNonlinearPrincipalComponent1991a},
allowing one to construct a lower dimensional latent ODE, which can be solved at, typically,
a lower cost \citep{grathwohlFFJORDFreeformContinuous2018, rubanovaLatentODEsIrregularlySampled}. This latent ODE can be defined by a small dummy chemical network \citep{grassiReducingComplexityChemical2021},
constant terms \citep{sulzerSpeedingAstrochemicalReaction2023} or a tensor expression akin
to a larger chemical network \citep{maesMACEMachineLearning2024}.

\subsection{The 3D-PDR Orion dataset}
The 3D-PDR code \citep{bisbas3DPDRNewThreedimensional2012} is a flexible code that can simulate photodissociation regions
in both 1D and 3D. For the purpose of this paper, we generated a dataset of 1D models,
each with a different initial condition of external radiation field \guvinit, density \densinit~
and cosmic ray ionisation \zetainit, inspired by the Orion cloud.
We call this physical parameter space  $P\in\mathbb{R}^3$. From this physical parameter space, we generate
a total of 8192 models using Sobol sampling; these models are then divided into a training, validation
and test set with a $0.70$, $0.15$ and $0.15$ split respectively. 

For each sample in $p \in P$, we obtain one series consisting of 215 relative abundances $x_i(\av)=n_i(\av)/\densinit$ as a function of 300 monotonously increasing visual extinctions (\av), illustrated in appendix \ref{avsteps}. 
This gives us a series $X_p\in \mathbb{R}^{300 \times 215 }$ for each point in the physical parameter space, $p \in P$, where each vector of abundances is linked to one visual ``depth''.
Besides the abundances of the molecules, 3D-PDR provides us with outputs such as the extinction \av, gas temperature \tgas(\av), dust temperature \tdust(\av), \guv, \densinit~and \zetainit, giving the  auxiliary
parameters $E_p \in \mathbb{R}^{300 \times 6}$, with the last two parameters constants.
We combine these aforementioned series into the dataset by picking the abundances of: \electron, \Hatom, \hydrogen, \Oatom, 
\CO, \water, \CH, \oxygen, \CN, \HCOp, \NH, \HCN, \carbon, \HCO, \formaldehyde, \COp, \CS~and \methanol, 
adding the auxiliary parameters $E_p$ and two of the constant physical parameters, namely \densinit and
\zetainit. This results in the complete dataset, with 8192 series of $D \in \mathbb{R}^{300 \times 25}$. 

In order to prevent extremely small values from creating a too large dynamic range
in log-space,
we add a minor offset to each of the features: $\epsilon=10^{-20}$ for the abundances $X_p$ and the auxiliary parameters $E_p$, and $\epsilon=10^{-10}$ for the visual extinction \av. Subsequently, we apply a $\log_{10}$ transformation.
Lastly, we standardize the data using the mean $\tilde{\mu}$ and standard deviation $\tilde{\sigma}$ for each of the log transformed features individually.
The data transformation combined is:
\begin{equation}
D'_i = \frac{\log_{10}(D_i+\epsilon_i) - \tilde{\mu}}{\tilde{\sigma}}
\end{equation}
% This provides us with all features in normalized log-space such that we can find an emulator that
% works well across the entire dynamic range of the simulation.

\subsection{On training neural differential equations}
In this paper, we investigate two different types of architectures.
Firstly the we explore the  direct application of ANODE on the data, mapping from $D_{i}$ directly to the
next visual extinction $D_{i+1}$,
hereafter called the \texttt{evolve (e)} model. The second architecture
uses an additional encoder and decoder, adapted from the encoder-decoder with bottleneck and latent ODE model from \cite{sulzerSpeedingAstrochemicalReaction2023}, which applies the same mapping but now with
the ANODE in the latent space, 
hereafter called the \texttt{encoder-evolve-decode (eed)} model.
To the latter, we also add the auxiliary parameters $E_p$ into the encoder with a latent bottleneck of size 5 
(\texttt{eed-a}), adding them in the latent space giving it size 5+4 (\texttt{eed-b}) and adding them into the encoder
but enlarging the bottleneck size to 9 (\texttt{eed-c}). All these models can be found at \url{https://github.com/uclchem/neuralpdr}.

In order to effectively train the evolve models, we use several 
methods. Firstly we use weight decay \citep{loshchilovDecoupledWeightDecay2017} to penalize 
large weights in the model, since they can lead to expensive
to evaluate and instable differential equations. Additionally
we initialize the weights for the ANODE with a truncated normal distribution and
scale it with a factor $\sqrt{b/n}$ \citep{heDeepResidualLearning2015a} with $b$ a hyperparameter
and $n$ the width of the inputs into the layer. This ensures the 
RHS of the differential starts out small and improves training.
Additionally, we use a learning
rate scheduler, namely a cosine delay schedule with a linear
warmup phase \citep{loshchilovSGDRStochasticGradient2016}.We also propose to introduce the
series in fractions, first training on the first part, then adding
the second parts and so forth. This is combined with the learning
rate scheduler, restarting the cosine delay schedule each time after introducing a new fraction. 

In summary for the direct models, we choose, after hyperparameter tuning, 
a neural network of 592 neurons wide,
3 layers deep, a softplus activation function between all layers, a peak learning rate of  $1.1\times10^{-3}$, a weight scale $b=1.3$ with no truncation between -10 and 10, a weight decay of $1.2\times10^{-4}$,  a batch size of 48 and the dataset split into three equal fractions, with
each fraction being expanded after 50 epochs, with the full dataset being trained on for an extra 50 epochs, resulting in 200 epochs in total. We then perform an ablation study
to compare the complete model with all these features, model (\texttt{e-a}),
to a model without the weight decay (\texttt{e-b}), a model without
the learning rate schedule (\texttt{e-c}), a model with normal
weight initialization (\texttt{e-d}), a model without introducing
the fractions of training data (\texttt{e-e}),
and lastly a model with none of the above (\texttt{e-f}).

\section{Experiments and results}

\begin{figure}
  \centering
  % \fbox{\rule[-.5cm]{0cm}{4cm} \rule[-.5cm]{4cm}{0cm}}
  \includegraphics[width=\linewidth]{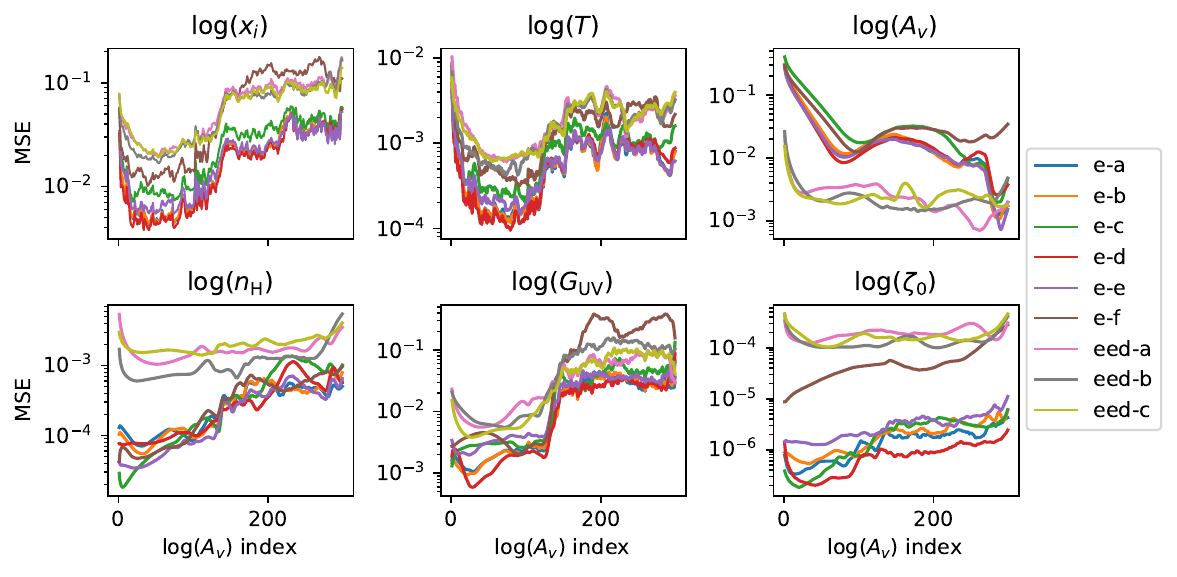}
  \caption{Comparison between different NeuralPDR configurations, the y-axis is the index
  of the visual extinction. The loss of the \texttt{a,b,c} and \texttt{e} \texttt{evolve} models largely overlap.}
  \label{modelcomparison}
\end{figure}

\begin{figure}
  \centering
  % \fbox{\rule[-.5cm]{0cm}{4cm} \rule[-.5cm]{4cm}{0cm}}
  \includegraphics[width=\linewidth]{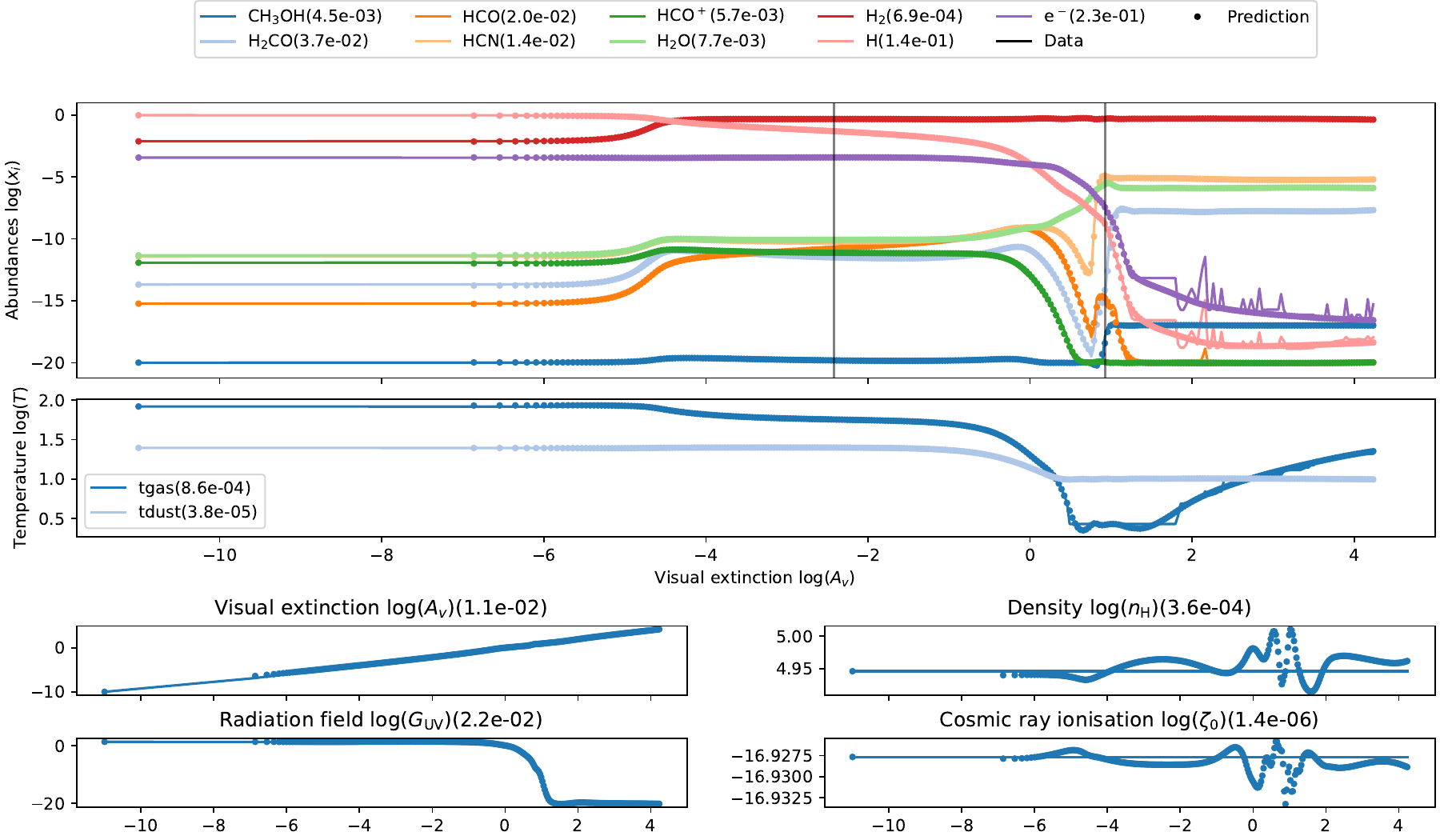}
  \caption{The result for one sample as inferred by the (\texttt{e-a}) emulator as a function of the log-\av. The MSE in the feature space for this sample is listed in parenthesis behind each feature. }
  \label{results:plot683}
\end{figure}

We find that both the encoder-evolve-decoder and the evolve models can emulate
3D-PDR in a satisfactory fashion. We firstly investigate the MSE over all
test samples per visual extinction index in figure \ref{modelcomparison}. This figure
shows that the first four evolve models (\texttt{e-a} through \texttt{e-d}) perform best in general, with every single
ablation not performing significantly worse, (\texttt{e-f}) with all features ablated,
shows a significantly worse performance. It is not clear which of the
improvements to the training process we proposed is exactly responsible
for the improvement in performance, but altogether they result in a well trained model.
The family of \texttt{eed} models does not perform as well as the \texttt{evolve} models for most features, which could be attributed to the 
small bottleneck size. Only for the \av~, the \texttt{eed} models do better,
which is counter-intuitive since the \av~ prediction is reused as an input feature for the next 
evolution of the model, but apparently so, the inaccurate \av~predictions are sufficient.

We show the predictions for a selection molecules, the temperatures and the auxiliary features 
inferred with the best model, (\texttt{e-a}) in figure \ref{results:plot683}, with all individual series in appendix \ref{allseriesplots}.
The model shows good agreement for most features, with the molecules at high abundances hard
to discern from the original data. Especially the high abundance, and therefore easily
observable by telescope, molecules show a good agreement between the data and the predictions.
This is also well within the underlying uncertainties of the underlying parameters 
as specified in the chemical databases \citep{millarUMISTDatabaseAstrochemistry2024}.
The molecules at low abundances have many fluctuations in
the data, the neuralODE however
fits a smooth function to it, providing a desirable smooth interpolation in this regime.
This is preferred to overfitting the jumps in the data since they are
artifacts of the 3D-PDR solver, caused by the iterative nature of solving
the temperature and chemistry balance. The temperature is fitted well,
with the gas temperature having a small deviation at the point where the trend in temperature changes rapidly. Both the changing and constant 
auxiliary features are reproduced well, with the constant features
having the lowest errors of all. 

The original dataset of 8192 samples was generated on an Intel(R) Core(TM) i9-13900 CPU system with 16 cores, taking 432 hours. The training
happens on a system with a NVIDIA RTX 2080, taking 2.5 hours for 200 epochs. Inference then takes 6 seconds for 1228 samples including loading, compiling and saving the data. 
This critical speed up is exactly what we need for simulating PDRs.

\section{Conclusions}
In this paper we introduced the 3D-PDR Orion dataset, which provides 8192 models that capture the chemistry of the Orion Bar. 
The emulators that we train on this dataset can be used for higher spatial and temporal resolution simulations, enabling us to resolve PDR regions in more detail.
We show that the benchmark ANODE architectures can provide high fidelity predictions at a fraction of the computational cost, especially with the addition of the weight decay, learning rate scheduler, small weight initialization and weight decay.
With these emulators, we enable fast computation
of the chemistry of PDR regions, either
by replacing the classical code altogether, or by using the emulator as a preconditioner for the classical code.
In future work, the dataset will be expanded upon to contain actual line traces of 3D simulations of different regions, with variable density, visual extinction and cosmic ray attenuation profiles along the line of sight \citep{gachesAstrochemicalImpactCosmic2019}.

\begin{ack}
G.V., R.R. and S.V. acknowledge support from the European Research Council (ERC) Advanced grant MOPPEX 833460.
T.G.B. acknowledges support from the Leading Innovation and Entrepreneurship Team of Zhejiang Province of China (Grant No. 2023R01008). The authors declare not competing interests. 

The ANODEs were implemented using \texttt{diffrax}\citep{kidgerNeuralDifferentialEquations2022} and \texttt{jax}\citep{jax2018github}. Plots were made using \texttt{matplotlib} \citep{Hunter:2007}. The dataset was serialized into its final format using \texttt{h5py} \citep{collettehdf5}.
\end{ack}

{
\small
\bibliographystyle{plainnat}
\interlinepenalty=10000
\bibliography{neuralpdr.bib}
}
%%%%%%%%%%%%%%%%%%%%%%%%%%%%%%%%%%%%%%%%%%%%%%%%%%%%%%%%%%%%

\appendix
\newpage
\section{The relation between \av~index and $\log(\av)$~ for different models}
\label{avsteps}
We show here the connection between the index of the visual extinction and its logs values.
\begin{figure}[ht]
  \centering
  % \fbox{\rule[-.5cm]{0cm}{4cm} \rule[-.5cm]{4cm}{0cm}}
  \includegraphics[width=\linewidth]{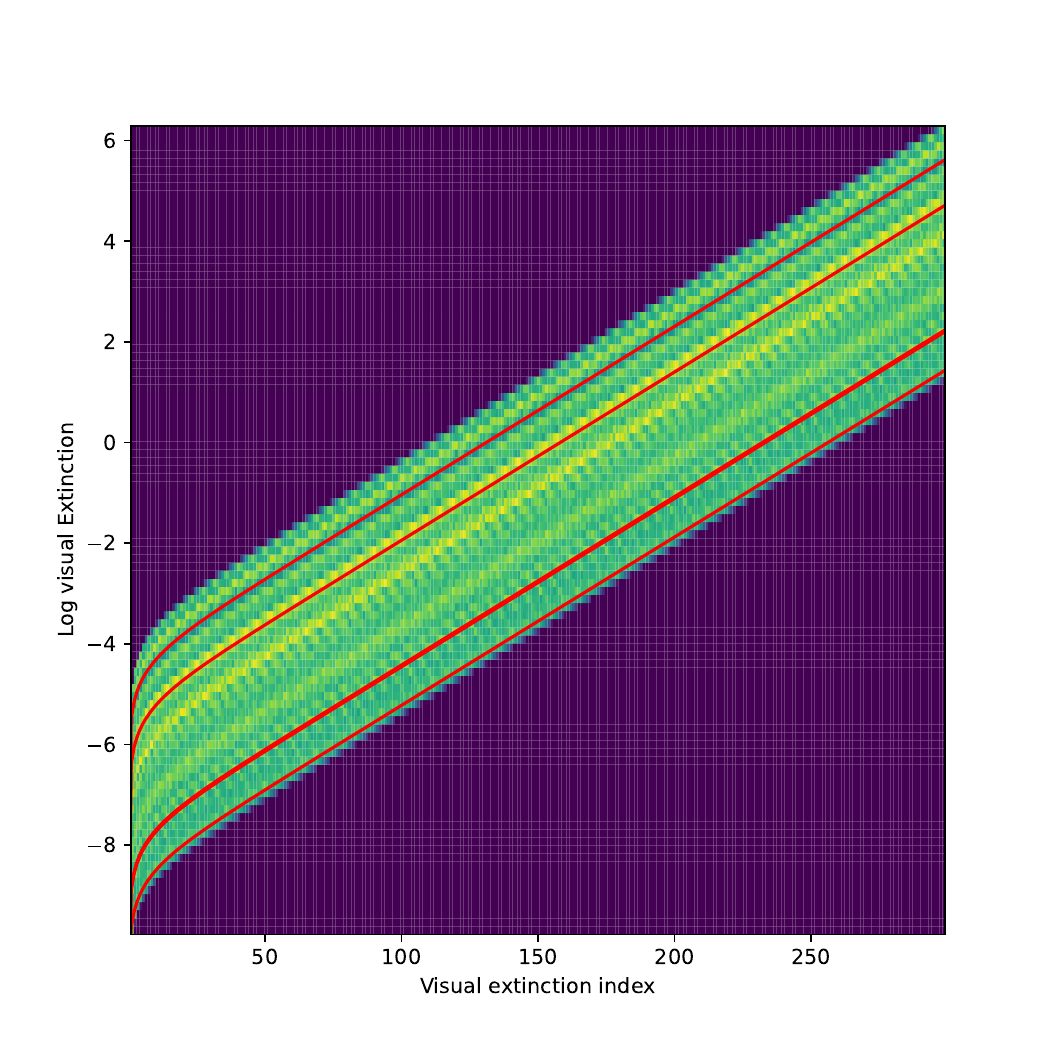}
  \label{results:avindexvsav}
  \caption{The distribution of visual extinction
  index versus the actual log values of the index. The 5 red lines are 5 individual samples, with the 2d histogram the distribution.}
\end{figure}
 \FloatBarrier

\newpage
\section{Loss function}
\label{valloss}
We provide here the validation loss as it is computed during the training process, which is directly on the standardized training output.
\begin{figure}[ht]
  \centering
  % \fbox{\rule[-.5cm]{0cm}{4cm} \rule[-.5cm]{4cm}{0cm}}
  \includegraphics[width=\linewidth]{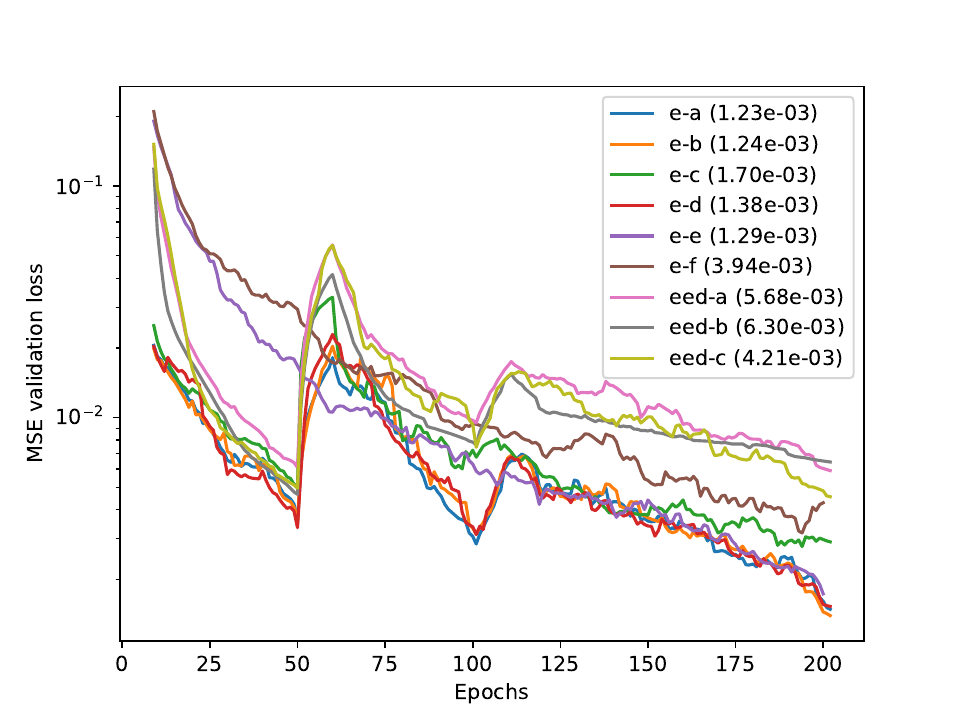}
  \label{results:allvalidation}
  \caption{A comparison of the validation for the losses, with a moving average of 10 steps.
  The final loss is denoted in the legend behind each architecture.}
\end{figure}
 \FloatBarrier
\newpage
\section{MSE for each individual feature}
The mean squared error in feature space for each of the series for the different models.
\label{allseriesplots}
\begin{figure}[ht]
  \centering
  % \fbox{\rule[-.5cm]{0cm}{4cm} \rule[-.5cm]{4cm}{0cm}}
  \includegraphics[width=0.7\linewidth]{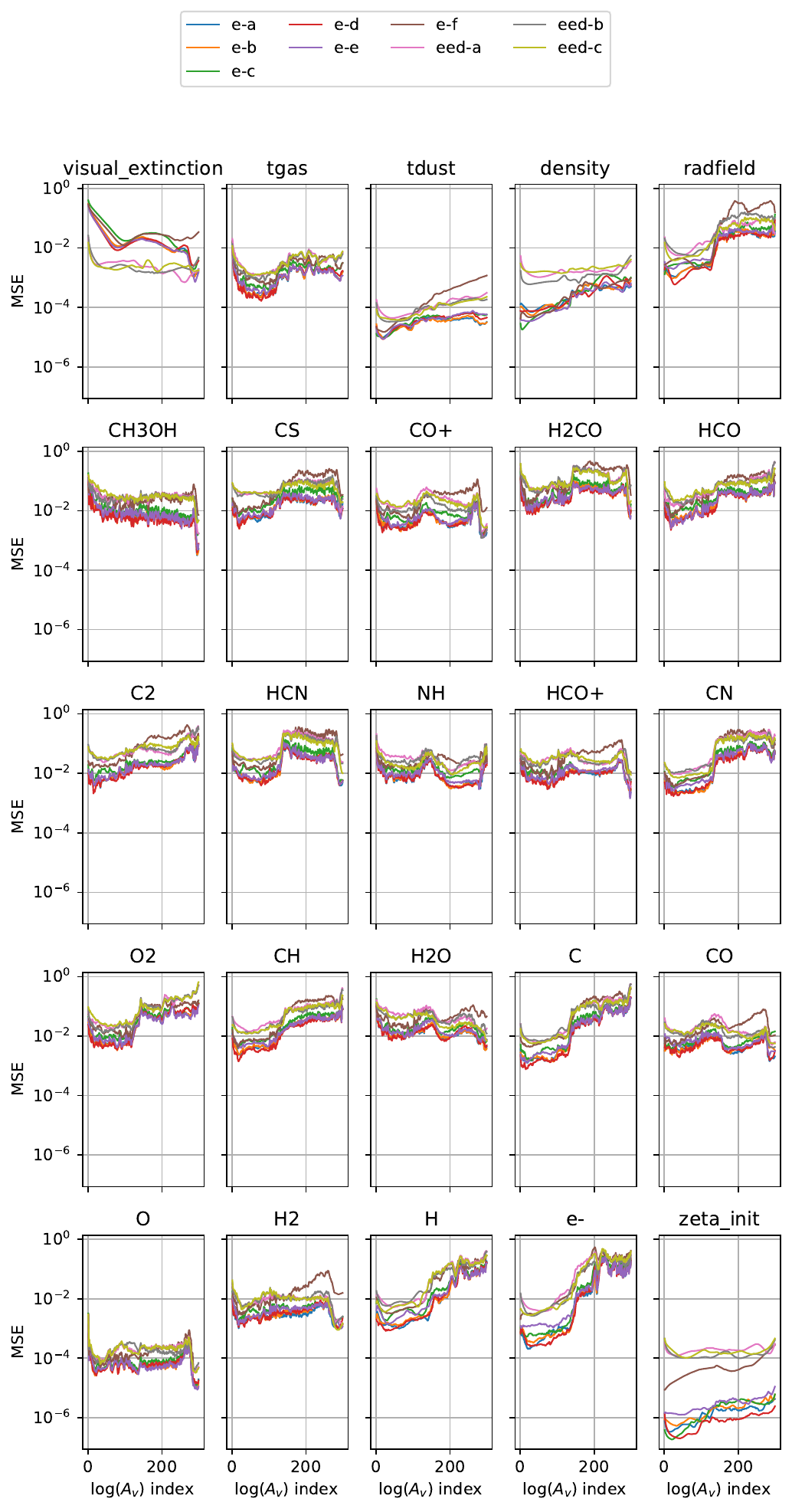}
  \label{results:allseriesmse}
  \caption{The MSEs for each individual molecule and all other features, for all different architectures.}
\end{figure}
 \FloatBarrier
\newpage

\end{document}